\documentclass[twocolumn,nofootinbib,prd]{revtex4}
\usepackage{amsmath}
\usepackage{graphicx}
\usepackage{amssymb}
\usepackage{mathrsfs}
\usepackage{graphtex}

\newcommand{\pt}{\partial_{t}}

\begin{document}

\title{Constant Mean Curvature Slices of the Reissner-Nordstr\"{o}m Spacetime}

\author{Patrick Tuite}
\email{p.tuite@ucc.ie}
\affiliation{Physics Department, University College Cork, Cork,
Ireland}

\author{Niall \'O~Murchadha}
\email{niall@ucc.ie}
\affiliation{Physics Department, University College Cork, Cork,
Ireland}

\begin{abstract}

\end{abstract}

%\keywords{}

\begin{abstract}
  In order to specify a foliation of spacetime by spacelike hypersurfaces we need to place some restriction on the initial data and from this derive a way
  to calculate the lapse function $\alpha$ which measures the proper time and interval between neighbouring
  hypersurfaces along the normal direction. Here we study a prescribed slicing known as
  the \textit{Constant Mean Curvature} (CMC) slicing. This slicing will be applied to the Reissner-Nordstr\"{o}m
  metric and the resultant slices will be investigated and then compared to those of the extended Schwarzschild
  solution.
\end{abstract}

\maketitle

\section{Introduction}
Analysing General Relativity in terms of a Hamiltonian system
\cite{Misner1973} a time function is chosen and one considers the
foliation of the spacetime by the slices of constant time. On these
spacelike slices two geometric quantities arise, the intrinsic
metric $g_{ij}$ and the extrinsic curvature $K^{ij}$. Both these
quantities are related to each other through
the constraints, in a vacuum,
\begin{eqnarray}
\mathcal{R}^{(3)} - K^{ij}K_{ij} + \left(\mathrm{tr}K\right)^2 & = &
0, \\
\nabla_{j}K^{ij} - g^{ij}\nabla_{j}\mathrm{tr}K & = & 0.
\end{eqnarray}
Where $\mathcal{R}^{(3)}$ is the three-scalar curvature.

Given initial data, an arbitrary lapse $\alpha$ and shift vector
$\beta^i$ can be chosen which will determine the magnitude and
direction of the unit time vector relative to the normal to the
spacelike slice. It is this choice, which can be seen both as a blessing and a curse,
as the question to be answered is, what constitutes a good or appropriate choice?

The evolution equations for the intrinsic metric and extrinsic
curvature are given by (note the convention here will follow
\cite{Wald} and not \cite{Misner1973})
\begin{eqnarray}
\pt g_{ij} & = & 2\alpha K_{ij} + \beta_{i;j} + \beta_{j;i}, \\
\pt K_{ij} & = & \alpha_{;ij} - \alpha\left(R_{ij} -
2K_{i}^{\;k}K_{kj} + K_{ij}K\right) \notag \\
& & + K_{ij;k}\beta^{k} +
K_{ik}\beta^{k}_{\; ;j} + K_{jk}\beta^{k}_{\;;i}.
\end{eqnarray}
Where $K = \mathrm{tr}K$. Using the convention of signs in
\cite{Wald} this gives $K = +n^{\alpha}_{\;;\alpha}$ where
$n^{\alpha}$ is the timelike unit normal to the slice and
$\pt\sqrt{g} = \sqrt{g}\left(\alpha K + \alpha^{i}_{\;;i}\right)$.
In this convention a positive $K$ gives expansion. A standard way to
choose a foliation, and thus time, is to place a condition on the
extrinsic curvature. Many different types of conditions are used
throughout the numerical relativity community, such as ``1+log''
slicing \cite{Bona1995a} or maximal slicing
\cite{Estabrook1973} where $K$ is chosen to be zero on each slicing.
Another popular slicing is to set the trace of the extrinsic
curvature to be constant on each slice, this is known as
\textit{constant mean curvature} slicing (`CMC slicing')
\cite{CMCpart1}.

Here we will investigate the CMC slices of the
Reissner-Nordstr\"{o}m spacetime. In \cite{Reimann2004, Reimann2004a}
maximal slicing of this metric has been analysed. The
Reissner-Nordstr\"{o}m metric is a spherically symmetric solution of
the coupled equations of Einstein and Maxwell. It represents a
stationary, non-rotating black hole of mass $m$ and a charge Q. The
metric for the Reissner-Nordstr\"{o}m can be written as
\begin{eqnarray}
g_{\mu\nu} & = & -\left(1 - \frac{2m}{r} +
\frac{Q^{2}}{r^2}\right)dt^{2} \notag \\
& & + \frac{dr^{2}}{\left(1 -
\frac{2m}{r} + \frac{Q^{2}}{r^2}\right)}+ r^2d\theta^2 +
r^2\sin{\theta}^{2}d\phi^2
\end{eqnarray}
in units where $G = c = 1/4\pi\epsilon_{0} = 1$ and where $t$ is the
`static' killing vector and $r$ is the areal radius.

CMC slices are of value to the numerical relativity community
dealing with the analysis of gravitational radiation. The waveforms
of the radiation are easier to detect on asymptotically null
surfaces which the CMC slices become as they approach null infinity
\cite{Rinne2010}.

The approach used to analyse the spacetime will follow a height function
approach used in \cite{CMCpart1}. Taking the Reissner-Nordstr\"{o}m spacetime and
performing a coordinate transformation in the $\left(t,r\right)$
plane, given by $t = h\left(r\right)$, leaving the other coordinates
unchained. $h\left(r\right)$ is called the height function. Now one
imposes the condition that the $t' = t - h\left(r\right) = 0$ slice be CMC. This condition
will produce a second order equation for the height function which can be
integrated explicitly once. From this the intrinsic metric and
extrinsic curvature can be obtained.

\section{CMC slicing}

From \cite{CMCpart1} there are two complementary methods to find and
analyse the CMC slices. As mentioned in the previous section there
is the height function approach using the coordinate transformation
$t = h\left(r\right)$ and the second method is using a general
spherically symmetric metric given by
\begin{eqnarray}
ds^2 = -\alpha^2dt^2 + adr^2 + R^2\left[d\theta^2 +
\sin\theta^2d\phi^2\right]
\end{eqnarray}
Here the geometry is encoded in two places. One is the dependence of
$a$ on $r$ and the other is on the relationship of $R$ (the areal radius) and $r$ (the chosen radial coordinate). This
second piece of information is contained in the mean curvature of
the surfaces of constant $r$ as embedded two-surfaces in the spatial
three geometry.

The analysis in this paper will use the height function approach to find the CMC
slices of the spacetime.

The metric is given by
\small{
\begin{eqnarray}
g_{\mu\nu} & = & \notag \\
& & \hspace{-1 cm} \left(\begin{array}{cccc}
  -\left(1 - \frac{2m}{r} +
\frac{Q^{2}}{r^2}\right) & 0 & 0 & 0 \\
  0 & \left(1 - \frac{2m}{r} +
\frac{Q^{2}}{r^2}\right)^{-1} & 0 & 0 \\
  0 & 0 & r^{2} & 0 \\
  0 & 0 & 0 & r^{2}\sin\theta^{2}
\end{array}\right). \notag \\
\end{eqnarray}}
Now introducing the coordinate transformation given by
\begin{subequations}
\begin{align}
t' & = t + h\left(r\right), \\
r' & = r, \\
\theta' & = \theta, \\
\phi' & = \phi.
\end{align}
\end{subequations}
And the transformation to the new coordinate system is given by
\begin{eqnarray}
g_{\mu\nu}' = \frac{\partial x^{\alpha}}{{\partial
x^{\mu}}^\prime}\frac{\partial x^{\beta}}{{\partial
x^{\nu}}^{\prime}} g_{\alpha\beta}.
\end{eqnarray}
The metric becomes
\begin{widetext}
\begin{eqnarray}
g_{\mu\nu}' = \left(\begin{array}{cccc}
                -\left(1 - \frac{2m}{r} +
\frac{Q^{2}}{r^2}\right) \hspace{0.65 cm} & -h^{\prime}\left(1 - \frac{2m}{r} +
\frac{Q^{2}}{r^2}\right) & 0 & 0\\
                -h^{\prime}\left(1 - \frac{2m}{r} +
\frac{Q^{2}}{r^2}\right) \hspace{0.65 cm} & -\left(h^{\prime}\right)^2\left(1 -
\frac{2m}{r} + \frac{Q^{2}}{r^2}\right) + \left(1 - \frac{2m}{r} +
\frac{Q^{2}}{r^2}\right)^{-1} & 0 & 0 \\
                0 & 0 & r^{2} & 0 \\
                0 & 0 & 0 & r^2\sin\theta^2
              \end{array}\right). \notag \\
\end{eqnarray}
\end{widetext}
and the inverse of this metric is
\begin{widetext}
\begin{eqnarray}
{g^{\mu\nu}}^{\prime} = \left(\begin{array}{cccc}
                      -\left(1 - \frac{2m}{r} +
\frac{Q^{2}}{r^2}\right)^{-1} + \left(1 - \frac{2m}{r} +
\frac{Q^{2}}{r^2}\right)\left(h^{\prime}\right)^2 \hspace{0.65 cm} &
-h^{\prime}\left(1 - \frac{2m}{r} +
\frac{Q^{2}}{r^2}\right) & 0 & 0 \\
                      -h^{\prime}\left(1 - \frac{2m}{r} +
\frac{Q^{2}}{r^2}\right) \hspace{0.65 cm} & \left(1 - \frac{2m}{r} +
\frac{Q^{2}}{r^2}\right) & 0 & 0 \\
                      0 & 0 & r^{-2} & 0 \\
                      0 & 0 & 0 & \left(r^2\sin\theta^2\right)^{-1}
                    \end{array}
\right). \notag \\
\end{eqnarray}
\end{widetext}
Where $h^{\prime} = \partial h/\partial r$. The intrinsic metric is
given by
\begin{eqnarray}
ds^2 & = & \left[-\left(h^{\prime}\right)^2\left(1 - \frac{2m}{r} +
\frac{Q^{2}}{r^2}\right) + \left(1 - \frac{2m}{r} +
\frac{Q^{2}}{r^2}\right)^{-1}\right]dr^2 \notag \\
& & + r^2\left(d\theta^2 +
\sin\theta^2\right).
\end{eqnarray}
The lapse $\alpha$ of the slice is given by
\begin{eqnarray}
\alpha & = & \frac{1}{\sqrt{{-g^{00}}}} \notag \\
 & = & \left[\left(1 - \frac{2m}{r} +
\frac{Q^{2}}{r^2}\right)^{-1} - \left(1 - \frac{2m}{r} +
\frac{Q^{2}}{r^2}\right)\left(h^{\prime}\right)^2\right]^{-\frac{1}{2}},\label{eqn:lapseh} \notag \\
\end{eqnarray}
the shift $\beta_{r}$ by
\begin{eqnarray}
\beta_{r} = \left[-h^{\prime}\left(1 - \frac{2m}{r} +
\frac{Q^{2}}{r^2}\right),0,0\right]. \label{eqn:betah}
\end{eqnarray}
The future pointing unit normal is given by
\begin{widetext}
\begin{eqnarray}
n^{\mu} = \frac{\left[\left(1 - \frac{2m}{r} +
\frac{Q^{2}}{r^2}\right)^{-1} - \left(1 - \frac{2m}{r} +
\frac{Q^{2}}{r^2}\right)\left(h^{\prime}\right)^2, h^{\prime}\left(1
- \frac{2m}{r} + \frac{Q^{2}}{r^2}\right),0,0\right]}{\sqrt{\left(1
- \frac{2m}{r} + \frac{Q^{2}}{r^2}\right)^{-1} - \left(1 -
\frac{2m}{r} + \frac{Q^{2}}{r^2}\right)\left(h^{\prime}\right)^2}}.
\notag \\
\end{eqnarray}
\end{widetext}
It can be seen that this future pointing normal is a time-like unit
normal since
\begin{eqnarray}
n_{\mu}n^{\mu} = -1.
\end{eqnarray}
The mean curvature of the $t' = 0$ is given by
\begin{eqnarray}
K & = & n^{\mu}_{;\mu} =
\frac{1}{\sqrt{-g}}\left(\sqrt{-g}n^{\mu}\right)_{,\mu},
\end{eqnarray}
where $g$ is the determinant of the transformed metric
\begin{eqnarray}
g = -r^4\sin\theta^2.
\end{eqnarray}
Using this and the future pointing unit normal the mean curvature of a slice defined by $t' = t - h = 0$ is
\begin{widetext}
\begin{eqnarray}
K =
\frac{1}{\sqrt{r^4\sin\theta^2}}\partial_{r}\left(\sqrt{r^4\sin\theta^2}\frac{h^{\prime}\left(1
- \frac{2m}{r} + \frac{Q^{2}}{r^2}\right)}{{\sqrt{\left(1 -
\frac{2m}{r} + \frac{Q^{2}}{r^2}\right)^{-1} - \left(1 -
\frac{2m}{r} +
\frac{Q^{2}}{r^2}\right)\left(h^{\prime}\right)^2}}}\right).\notag \\
\label{eqn:meancurvature}
\end{eqnarray}
\end{widetext}
Since the chosen slicing is a constant mean curvature slicing,
$K$ is a constant over the whole $t' = 0$ slice.
Therefore we can integrate (\ref{eqn:meancurvature}) easily to give
\begin{eqnarray}
\frac{K r}{3} - \frac{C}{r^2} = \frac{h^{\prime}\left(1 -
\frac{2m}{r} + \frac{Q^{2}}{r^2}\right)}{{\sqrt{\left(1 -
\frac{2m}{r} + \frac{Q^{2}}{r^2}\right)^{-1} - \left(1 -
\frac{2m}{r} + \frac{Q^{2}}{r^2}\right)\left(h^{\prime}\right)^2}}}.\notag \\
\end{eqnarray}
Where C is a constant of integration. This can now be manipulated to
give an expression for $h'$
\begin{eqnarray}
h' = \frac{\frac{K
r}{3}-\frac{C}{r^2}}{\left(1-\frac{2m}{r}+\frac{Q^2}{r^2}\right)\sqrt{\left(
1 -\frac{2m}{r}+\frac{Q^2}{r^2}\right)
 +\left(\frac{K r}{3}-\frac{C}{r^2}\right)^2}}. \notag \\
\end{eqnarray}
Using this expression for $h'$ and substituting into
(\ref{eqn:lapseh}) and (\ref{eqn:betah}) will give very useful
expressions for the lapse and shift on the $t' = 0$ slice
\begin{eqnarray}
\alpha = \sqrt{\left( 1 -\frac{2m}{r}+\frac{Q^2}{r^2}\right)
+\left(\frac{K r}{3}-\frac{C}{r^2}\right)^2}, \label{eqn:lapsefinal}
\end{eqnarray}
and
\begin{eqnarray}
\beta_r = \frac{-\frac{K r}{3}+\frac{C}{r^2}}{\sqrt{\left( 1
-\frac{2m}{r}+\frac{Q^2}{r^2}\right)
 +\left(\frac{K r}{3}-\frac{C}{r^2}\right)^2}}. \label{eqn:betafinal}
\end{eqnarray}
Using these expressions for the lapse (\ref{eqn:lapsefinal}),
shift (\ref{eqn:betafinal}) and using the evolution equation for the
extrinsic curvature given by
\begin{eqnarray}
2\alpha K_{ab} = \frac{\partial g_{ab}}{\partial \bar{t}} -
\beta_{a;b} - \beta_{b;a}.
\end{eqnarray}
The term $\frac{\partial g}{\partial \bar{t}}$ in this case is zero
as $\left(\alpha, \beta_r\right)$ are the lapse and shift of the timelike Killing vector, since
$\alpha^2 - \beta^2 = \left( 1
-\frac{2m}{r}+\frac{Q^2}{r^2}\right)$. This gives for the extrinsic
curvature
\begin{eqnarray}
K^{r}_{\;r} & = & \frac{K}{3} + \frac{2C}{r^3} \\
K^{\theta}_{\;\theta} & = & \frac{K}{3} - \frac{C}{r^3} \\
K^{\phi}_{\;\phi} & = & \frac{K}{3} - \frac{C}{r^3}
\end{eqnarray}
This can be recognised as a combination of the trace term plus a
unique (up to a constant) spherically symmetric TT tensor, the terms
with coefficient $C$. Therefore these CMC slices of the
Reissner-Nordstr\"{o}m spacetime are completely defined by the two
parameters $K$ and $C$. This is the exact same result as for the
Schwarzschild solution \cite{CMCpart1}. The next step is to
see if further analysis of the system will correspond to the
analysis in \cite{CMCpart1}.

\section{Cylindrical CMC Slices}

Now we want to investigate cylindrical slices of this space-time. In
the upper $r < 2m $ and lower $r > 2m$ regions, the killing vector is
space-like and runs along the r = constant surfaces. Since
everything is constant along the Killing vector, the trace of the
extrinsic is preserved along these cylindrical surfaces. Therefore,
each r = constant surface is a CMC slice.

The new line segment of three metric can be written as
\begin{eqnarray}
dS^2 = \frac{dr^2}{1 - \frac{2m}{r} + \frac{Q^2}{r^2} +
\left(\frac{K r}{3} - \frac{C}{r^2}\right)^2} + R^2d\Omega^2.
\end{eqnarray}
The $dr^2$ term plays an important role in the calculation. Let
\begin{eqnarray}
k^2 = {1 - \frac{2m}{r} + \frac{Q^2}{r^2} + \left(\frac{K r}{3} -
\frac{C}{r^2}\right)^2}.\label{eqn:little k}
\end{eqnarray}
We see that $k = dr/dL$ where $L$ is the proper
length along the slice, and $2k/r$ is the 2-mean curvature of the
round 2-spheres as embedded in the 3-slice \cite{CMCpart1}. Also it can be seen from
(\ref{eqn:lapsefinal}) that $k = \alpha$ where $\alpha$ is the
killing lapse, is the dot product of the timelike Killing
vector with the unit normal to the slice.

\begin{figure}
  \centering
  \includegraphics[width=8 cm]{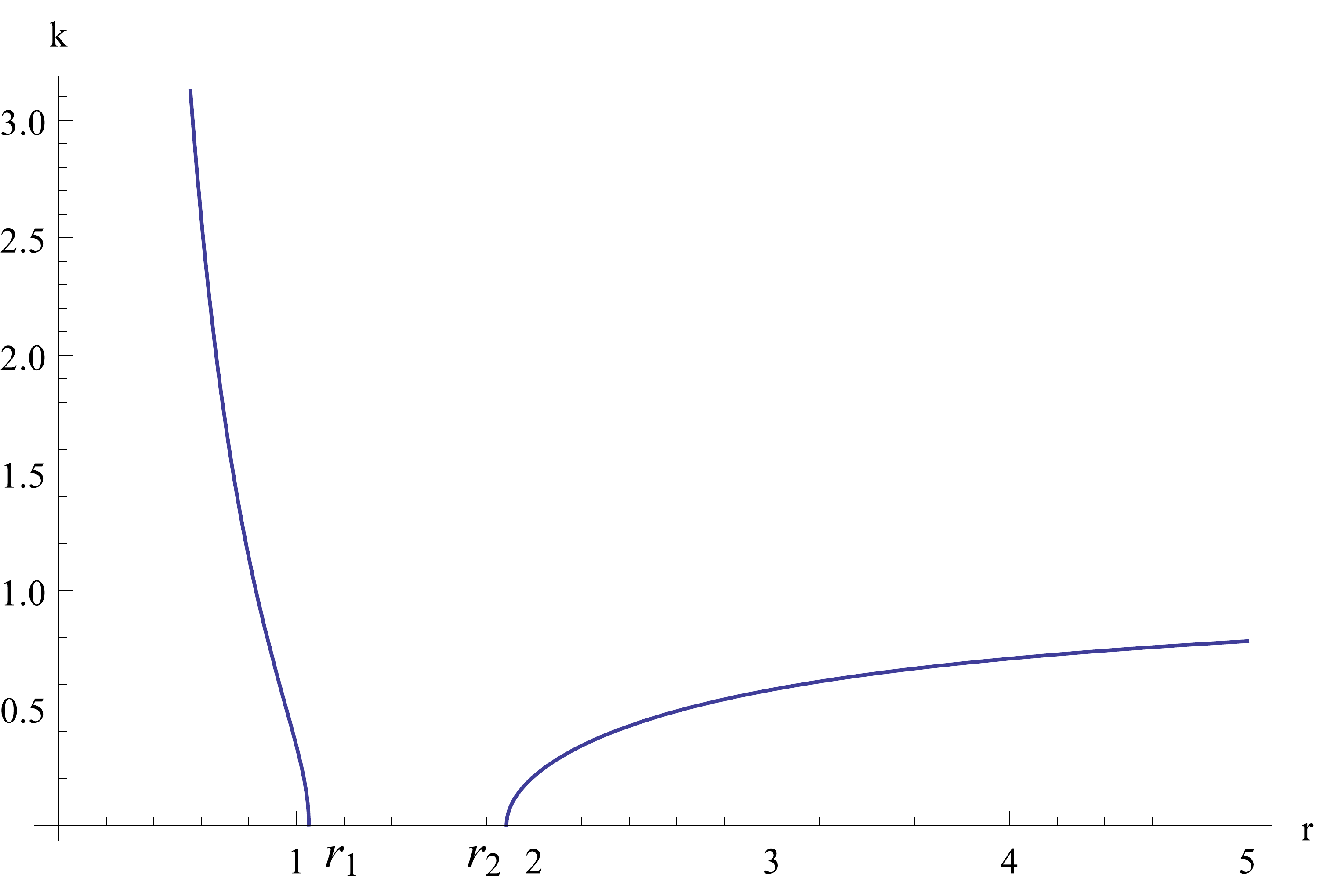}\\
  \includegraphics[width=8 cm]{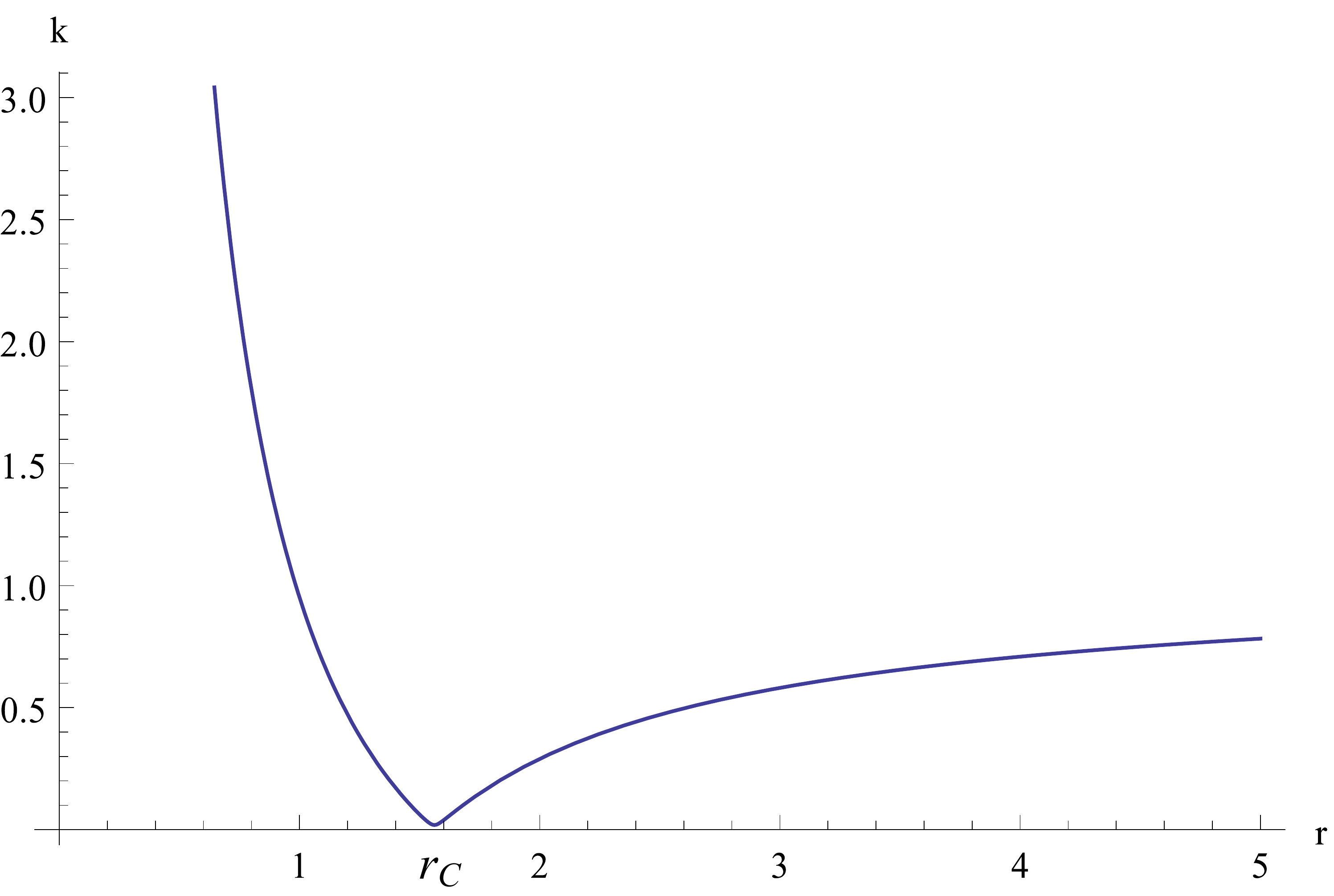}\\
  \includegraphics[width=8 cm]{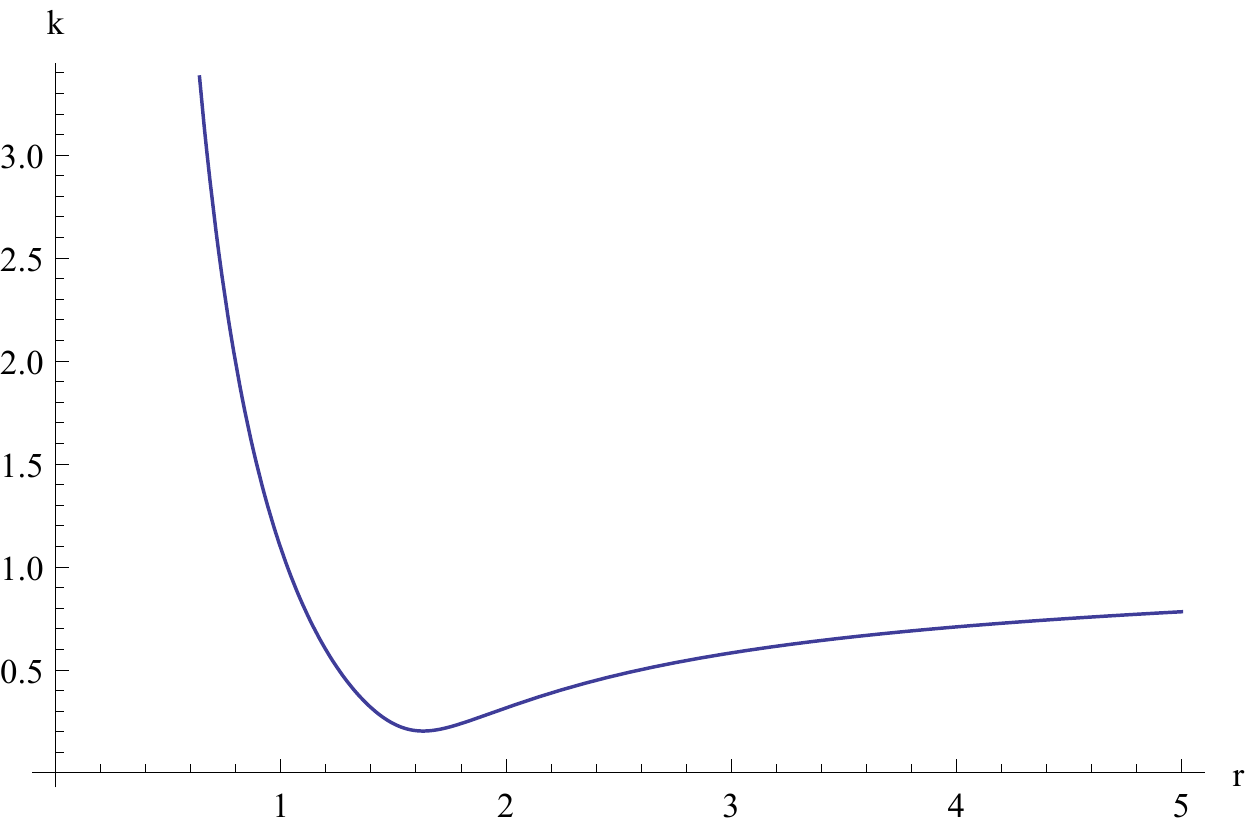}\\
  \caption{Profile curves of $k$. In the first panel both $K$ and $C$ are small and fixed. In the second and third panel the value of $K$ is held fixed and the value of $C$ is increased.}\label{fig:3sol}
\end{figure}

\begin{figure}
  \centering
  \includegraphics[width=6.5 cm]{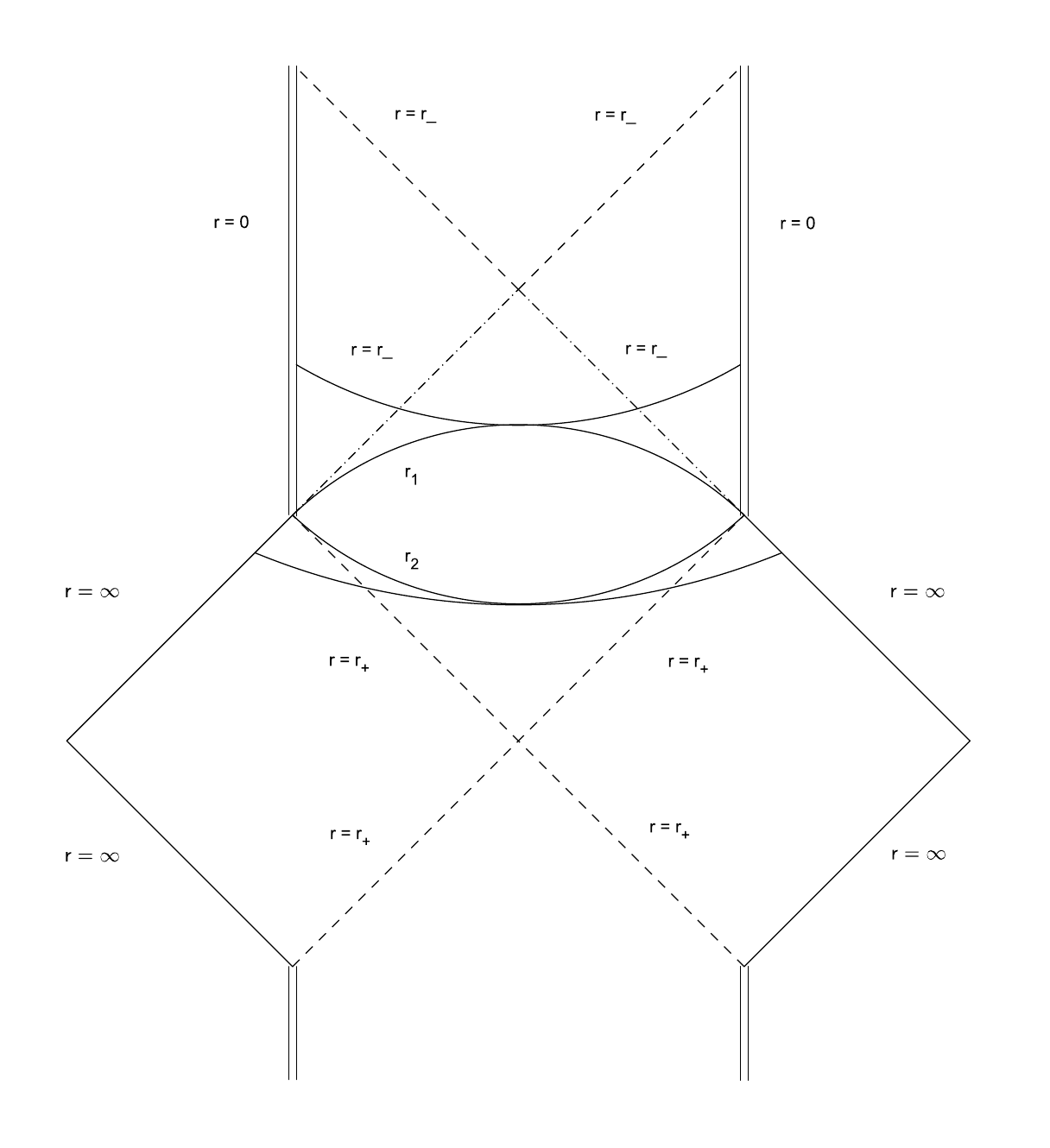}\\
  \includegraphics[width=6.5 cm]{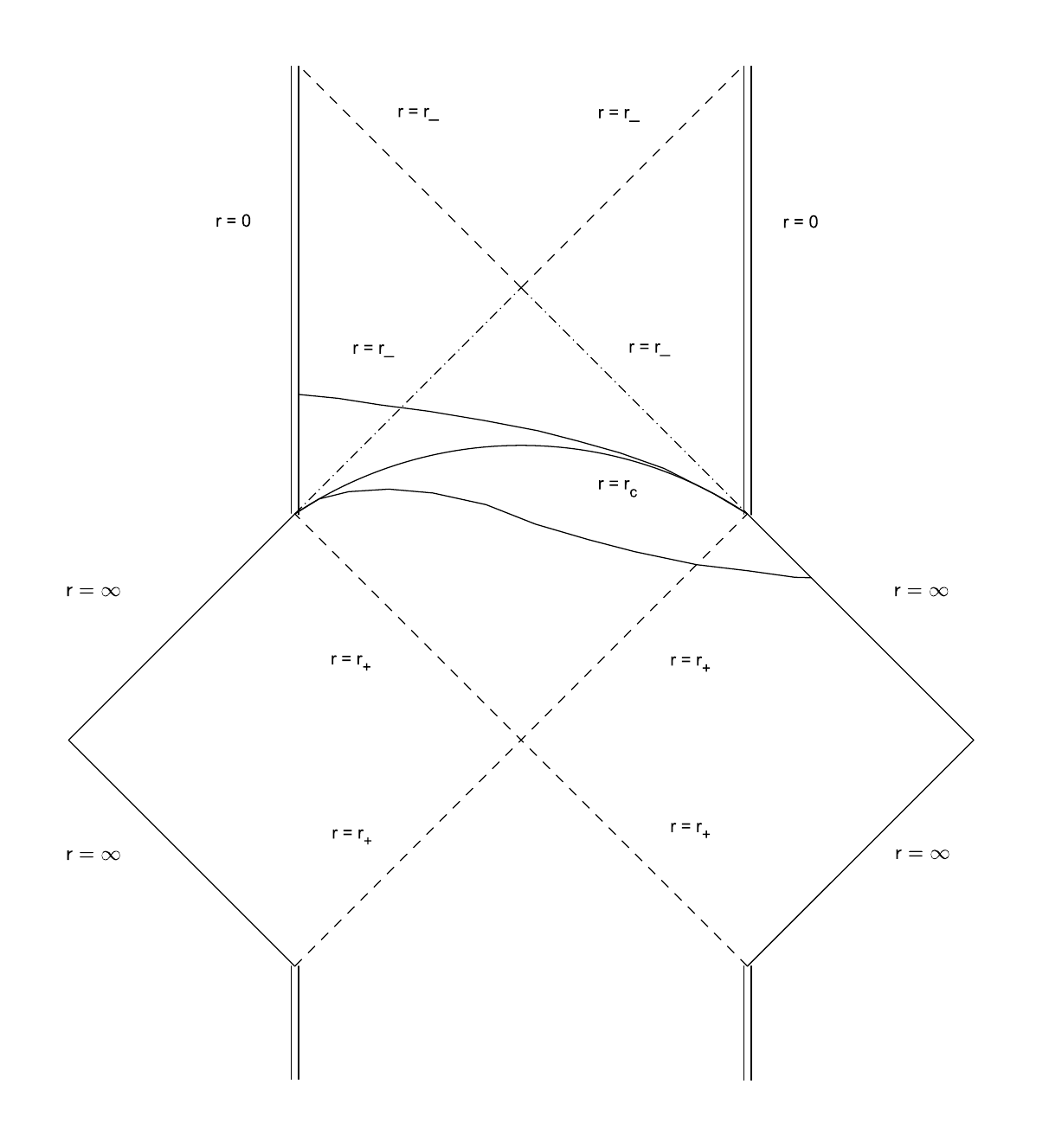}\\
  \includegraphics[width=6.5 cm]{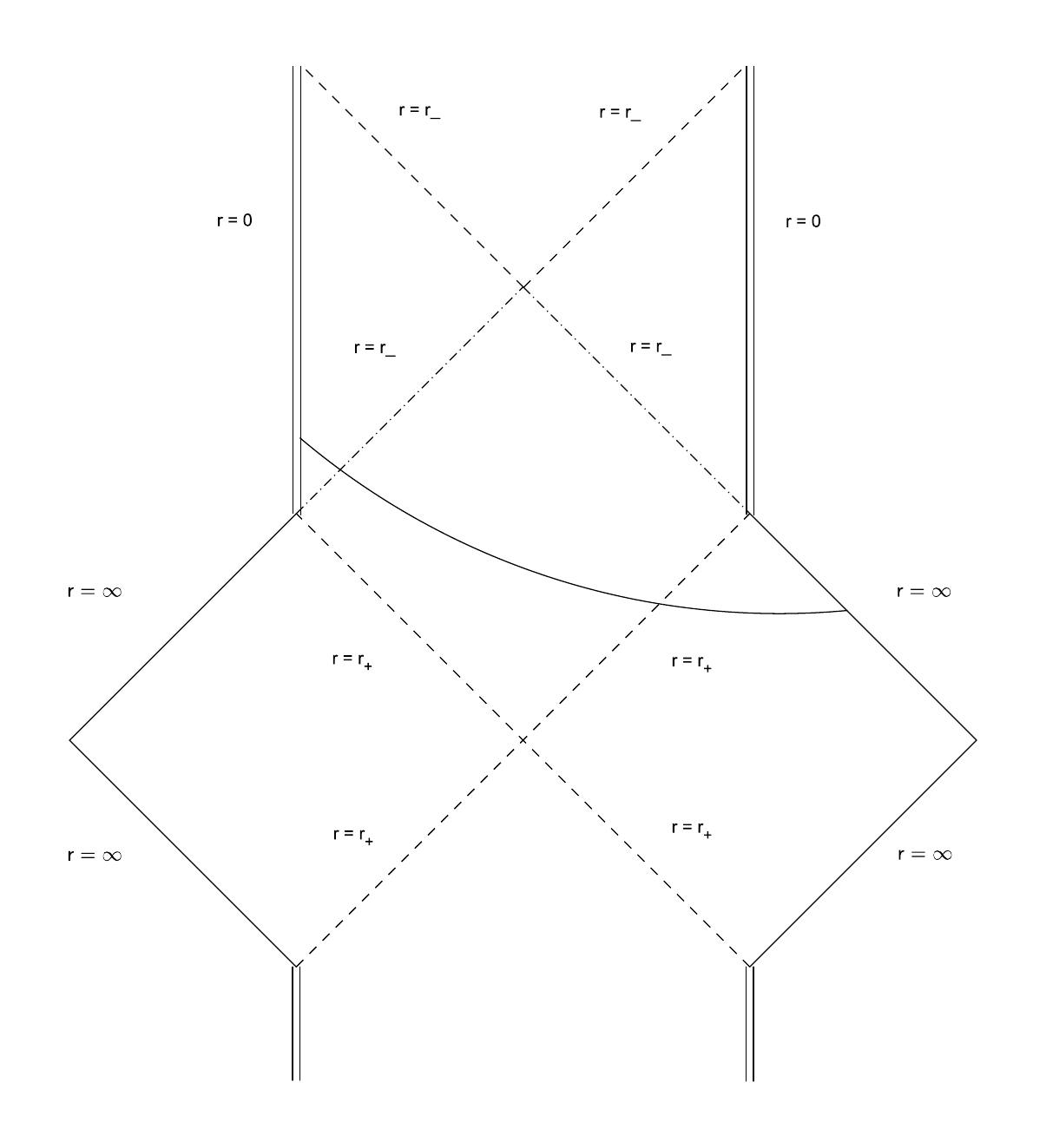}\\
  \caption{Penrose diagrams to illustrate the behaviour of $k$ as the value of $C$ is increased. In the first panel $(K, C)$ are fixed. Panel 2 shows the behaviour at the critical value for $C$. The last panel shows the behaviour as $C$ is increased beyond the critical value.}\label{fig:Penrose}
\end{figure}

Useful information on the behaviour of $k$ can be extracted by examining the profile of $k$ and the roots of the equation $k = 0$.
Fixing values for $\left(m, Q < m\right)$ and choosing 'small' values for $K$ and $C$. From the first panel of figure \ref{fig:3sol} we can see that there are 2 solution curves since $k$ cannot be less than zero. One of the solution curves starts out at the singularity $(r = 0)$ and approaches $r = r_1$. Here it bounces (since $k = 0, r = r_1$ is a maximal 2-surface) and heads back to $r = 0$. The other curve starts from null infinity and approaches the $r = r_2$ curve and again returns to null infinity. Examining the corresponding Penrose diagram (figure \ref{fig:Penrose} panel 1) allows us to see the behaviour of the CMC slices in detail.

Holding $K$ fixed and now increasing the value of $C$ we can see that the behaviour of $k$ will go through some interesting changes . Increasing the value of $C$ we can see that a transition occurs and a minimum in the profile of $k$ appears, panel 2 figure \ref{fig:3sol}. At this value of $C$ $r_1 = r_2 = r_C$. At this critical value of $C$ we have 3 possible curves (figure \ref{fig:Penrose} panel 2), firstly there is the curve starting at the black hole $(r = 0)$ and approaches $r_C$. The second curve is the cylindrical slice which occurs at $r_C$. And the final curve is the curve which starts at null infinity and approaches $r_C$. What we can see as the value of $C$ is increased and reaches this critical value trumpet slices \cite{Hannam2008} are formed.

Increasing the value of $C$ further beyond this critical value of $C$ there ceases to be any real solutions of $k = 0$, panel 3 figure \ref{fig:3sol}. Here the slices start at null infinity and travel to the black hole (figure \ref{fig:Penrose} panel 3).

\begin{figure}
  \centering
  \includegraphics[width=8 cm]{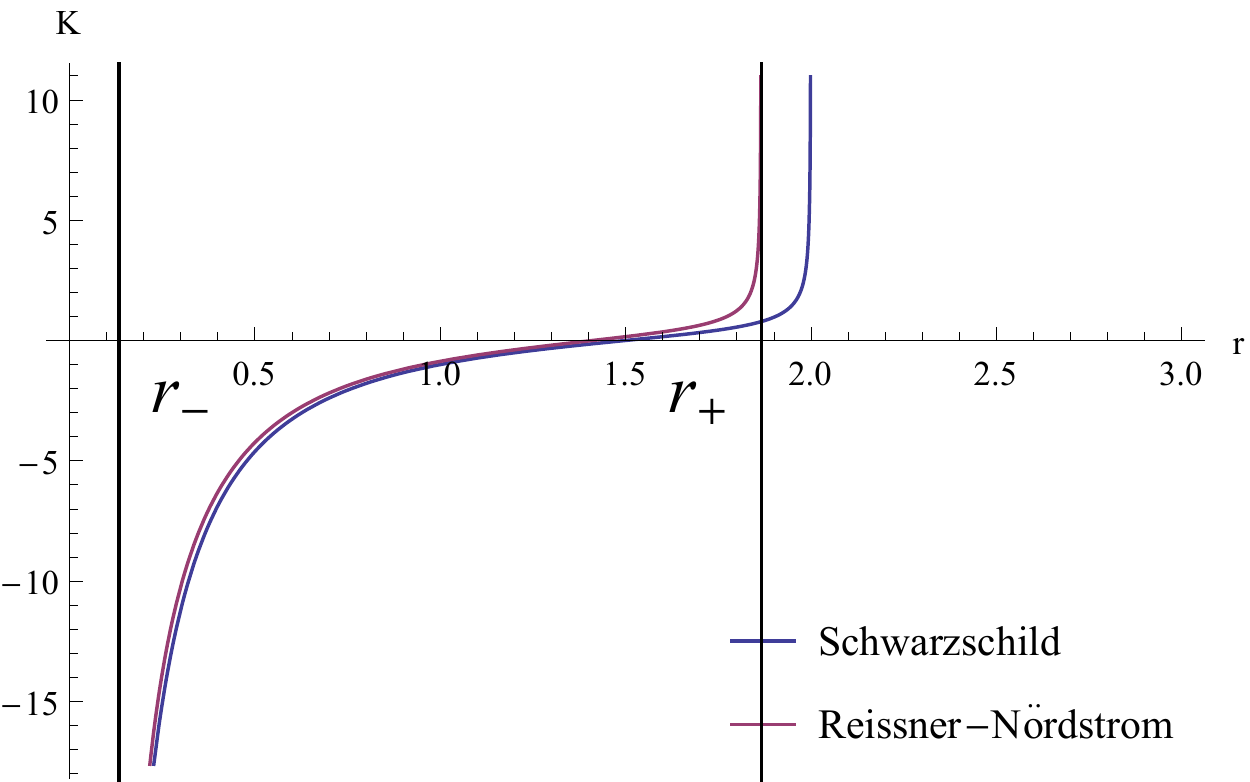}\\
  \caption{Graph of $K$ versus $r$ for $Q = 0.5$ for both the Schwarzschild \cite{CMCpart1} and Reissner-Nordstr\"{o}m expressions for $K$}\label{fig:kvr}
\end{figure}

The trace of the extrinsic curvature is given by
\begin{eqnarray}K & = & n^{\mu}_{;\mu} =
\frac{1}{\sqrt{-g}}\left(\sqrt{-g}n^{\mu}\right)_{,\mu}.
\end{eqnarray}
Here the normal vector $n^{\mu}$ is the normal to the r = constant
surface. This is given by
\begin{eqnarray}
n_{const r}^{\mu} = \left(0,\sqrt{1 - \frac{2m}{r} +
\frac{Q^2}{r^2}+\left(\frac{K r}{3} -
\frac{C}{r^2}\right)^2},0,0\right).
\end{eqnarray}
So the trace of the extrinsic curvature (in the upper quadrant) is
given by
\begin{eqnarray}
K = \frac{Q^2 + r\left(2r - 3m\right)}{\sqrt{Q^2r^4 - 2mr^5 + r^6}}. \label{eqn:Kafterh}
\end{eqnarray}
Since in the upper quadrant $r < 2m$ this can be written
\begin{eqnarray}
K = \frac{Q^2 + r\left(2r - 3m\right)}{\sqrt{-Q^2r^4 + 2mr^5 - r^6}}.
\end{eqnarray}
We can see from figure \ref{fig:kvr}, that K is large and negative near $r = r_-$, increases monotonically towards $r = r_+$ where it is large and positive and is zero at $r = \frac{1}{4}\left(3m \pm \sqrt{9m^2 - 8Q^2}\right)$. Comparing (\ref{eqn:Kafterh}) to the expression obtained in \cite{CMCpart1} for $K$ we can see from figure \ref{fig:kvr} the behaviour of $K$ is very similar to that of the Schwarzschild $K$.

The quadratic form $r^2 - 2mr + Q^2$ plays a key role. If $Q < m$, it has 2 real roots $0 < r_- < r_+ < 2m$. Only when
\begin{eqnarray}
1 - \frac{2m}{r} + \frac{Q^2}{r^2} < 0,
\end{eqnarray}
i.e. only when $r_- < r < r_+$ can $k^2$ be zero. Of course the surfaces satisfying $k = 0$ and $\frac{\partial k}{\partial r}$ are the cylindrical slices.

Given $r$, an expression for $C$ can be calculated using using $k$
above (\ref{eqn:little k}) and setting this equal to zero. This is
possible as we are effectively choosing a value of $r$ to give $k =
0$, therefore the corresponding $C$ is given by
\begin{eqnarray}
C = \frac{r^2 \left(2 Q^2+r (-3 m+r)\right)}{3 \sqrt{-r^2
\left(Q^2+r (-2 m+r)\right)}}.
\end{eqnarray}

In the lower quadrant $K$ becomes
\begin{eqnarray}
K = \frac{3 m r-2 r^2 -Q^2}{\sqrt{-Q^2r^4 + 2mr^5 - r^4}},
\end{eqnarray}
and similarly $C$ becomes
\begin{eqnarray}
C = \frac{r^2 \left(2 -Q^2+r (+3 m-r)\right)}{3 \sqrt{-r^2
\left(Q^2+r (-2 m+r)\right)}}.
\end{eqnarray}
Dropping the $Q^2$ terms these expressions reduce to the ones
obtained for the cylindrical CMC slices of the Schwarzschild metric.
From these results we can conclude that the many of the numerical
approaches which are applied to the CMC slicing of the Schwarzschild
metric can also be applied to the Reissner-Nordstr\"{o}m metric.
This has great benefits as the Reissner-Nordstr\"{o}m metric can act
as an analogy for a charged dust metric used in cosmology
\cite{Plebanski2006}. This allows for cross area comparisons of
numerical methods and results. And knowing the results can be
compared to well know results for Schwarzschild can add confidence
to the output of simulations.

\section{Generalised Lapse Function}

Following on from the work in \cite{CMCpart1} Malec and
\'O Murchadha found general spherically symmetric constant mean
curvature foliations of the Schwarzschild solution \cite{Malec2009}.
In this paper a generalised lapse function was found for the CMC
slicings of the Schwarzschild metric. The Einstein equations give an
evolution equation for the trace of the extrinsic curvature $K$, in
vacuum
\begin{eqnarray}
\nabla^2 \alpha - K^{ij}K_{ij}\alpha = \frac{\partial K}{\partial t}
- \beta^i\partial_{i}K.
\end{eqnarray}
Since we are taking $K$ to be a spatial constant this reduces to
\begin{eqnarray}
\nabla^2\alpha - K^{ij}K_{ij}\alpha = \frac{\partial K}{\partial t}.
\label{eqn:elliptic cmc alpha}
\end{eqnarray}
This is an equation for the Lapse function of a CMC slicing. It is
an elliptic equation which satisfies the maximum principle. In
\cite{Malec2009} it is found that the $G_{RR} = 0$ Einstein equation
can be written as
\begin{eqnarray}
\pt\left(R K^{\theta}_{\;\theta}\right)|_{R = \mathrm{const}} =
k^3\partial_{R}\frac{\alpha}{k}.
\end{eqnarray}
In this paper the above equation was solved using a Lapse function
\begin{eqnarray}
\alpha \equiv \delta k + k\int^{\infty}_{R}dr \frac{\dot{C} -
\frac{r^3}{3}\dot{K}}{r^2 k^3}. \label{eqn:lapse solution}
\end{eqnarray}
Where $\delta$ is a time dependent constant, $\dot{C} =\pt C$ and
$\dot{K} = \pt K$. In order to study a greater number of CMC
slicings both $K$ and $C$ were both parameterised using a time
parameter $t$. This parameterisation is not unique since one could
change the parameter and change the form of $C(t)$ and $K(t)$ but
the ratio of $\pt K/\pt C$ would not change. Inserting
(\ref{eqn:lapse solution}) into (\ref{eqn:elliptic cmc alpha}) it is
possible to verify that it is a valid solution for the lapse.

Since the results of the analysis CMC slicing of the
Reissner-Nordstr\"{o}m metric reduce to those of the Schwarzschild
solution it would be natural to assume that altering the Lapse
function in (\ref{eqn:lapse solution}) to correspond to that of the
Reissner-Nordstr\"{o}m slicing would satisfy (\ref{eqn:elliptic cmc
alpha}) also.

Taking equation (\ref{eqn:elliptic cmc alpha}) and performing the
differentiations
\begin{eqnarray}
\nabla^2\alpha - K^{ij}K_{ij} & = & \frac{\partial K}{\partial t},
\notag \\
g^{ij}\nabla_{j}\nabla_{i}\alpha - K^{ij}K_{ij}\alpha & = &
\frac{\partial K}{\partial t}, \notag \\
g^{ij}\nabla_{j}\zeta_{i} - - K^{ij}K_{ij}\alpha & = &
\frac{\partial K}{\partial t}.
\end{eqnarray}
Where we have used the fact that $\alpha$ is a scalar function to
write
\begin{eqnarray}
\partial_{i}\alpha = \zeta_{i}.
\end{eqnarray}
Taking the first term in the equation above
\begin{eqnarray}
g^{ij}\nabla_{j}\zeta_i = g^{ij}\left(\partial_{j}\zeta -
\Gamma^{l}_{ij}\zeta_{l}\right).
\end{eqnarray}
Reversing the substitution gives
\begin{eqnarray}
g^{ij}\nabla_{j}\partial_{i}\alpha & = &
g^{ij}\left(\partial_{j}\partial_{i}\alpha -
\Gamma^{l}_{ij}\partial_{l}\alpha\right), \notag \\
& = & g^{ij}\partial_{j}\partial_{i}\alpha -
g^{ij}\Gamma^{l}_{ij}\partial_{l}\alpha.
\end{eqnarray}
Since $\alpha$ is a function of $r$ only
\begin{eqnarray}
g^{ij}\nabla_{j}\partial_{i}\alpha & = &
g^{rj}\partial_{j}\partial_{r}\alpha -
g^{ij}\Gamma^{r}_{ij}\partial_{r}\alpha.
\end{eqnarray}
So the only contributing Christoffel symbols are
\begin{eqnarray}
\Gamma^{r}_{rr} & = & -\frac{\partial_{r}k}{k^2}, \\
\Gamma^{r}_{\theta\theta} & = & -r k, \\
\Gamma^{r}_{\phi\phi} & = & -r k \sin^2\theta.
\end{eqnarray}
Inserting these expressions into the above equation gives
\begin{eqnarray}
g^{ij}\nabla_{j}\partial_{i}\alpha & = &
g^{rj}\partial_{j}\partial_{r}\alpha -
\left(g^{rr}\left(-\frac{\partial_{r}k}{k}\right) \right. \notag \\
& & \left.+
g^{\theta\theta}\left(-r k\right) + g^{\phi\phi}\left(-r k
\sin^2\theta\right)\right)\partial_{r}\alpha.
\end{eqnarray}
Inserting the values for the metric components yields
\begin{eqnarray}
g^{ij}\nabla_{j}\partial_{i}\alpha & = &
g^{rj}\partial_{j}\partial_{r}\alpha - \left(-\partial_{r}k\ +
\left(-\frac{k}{r}\right) + \left(-\frac{k}{r}
\right)\right)\partial_{r}\alpha \notag \\
& = & g^{rj}\partial_{j}\partial_{r}\alpha - \left(-\partial_{r}k\ -
2\frac{k}{r}
\right)\partial_{r}\alpha \notag \\
& = & g^{rj}\partial_{j}\partial_{r}\alpha + \left(\partial_{r}k +
2\frac{k}{r} \right)\partial_{r}\alpha
\end{eqnarray}
Since the metric is diagonal the first term in the above equation
becomes
\begin{eqnarray}
g^{ij}\nabla_{j}\partial_{i}\alpha & = &
g^{rr}\partial_{r}\partial_{r}\alpha + \left(\partial_{r}k +
2\frac{k}{r}
\right)\partial_{r}\alpha \notag \\
& = & k\partial_{r}\partial_{r}\alpha + \left(\partial_{r}k +
2\frac{k}{r} \right)\partial_{r}\alpha.
\end{eqnarray}
Inserting (\ref{eqn:lapse solution}) into the equation gives
\begin{widetext}
\begin{eqnarray}
g^{ij}\nabla_{j}\partial_{i}\alpha = k\partial_{r}\partial_{r}\left(\delta k + \int^{\infty}_{R}dr
\frac{\dot{C} - \frac{r^3}{3}\dot{K}}{r^2 k^3}\right) +
\left(\partial_{r}k + 2\frac{k}{r} \right)\partial_{r}\left(\delta k
+ \int^{\infty}_{R}dr
\frac{\dot{C} - \frac{r^3}{3}\dot{K}}{r^2 k^3}\right). \notag \\
\end{eqnarray}
(\ref{eqn:elliptic cmc alpha}) becomes
\begin{eqnarray}
& & k\partial_{r}\partial_{r}\left(\delta k + k\int^{\infty}_{R}dr
\frac{\dot{C} - \frac{r^3}{3}\dot{K}}{r^2 k^3}\right) +
\left(\partial_{r}k + 2\frac{k}{r} \right)\partial_{r}\left(\delta k
+ k\int^{\infty}_{R}dr
\frac{\dot{C} - \frac{r^3}{3}\dot{K}}{r^2 k^3}\right) \notag \\
& & - K^{ij}K_{ij}\left(\delta k + k\int^{\infty}_{R}dr
\frac{\dot{C} - \frac{r^3}{3}\dot{K}}{r^2 k^3}\right) =
\frac{\partial K}{\partial t}.
\end{eqnarray}
The only non derivative terms which need to be evaluated now are the
$K^{ij}K_{ij}$. The diagonal components of the extrinsic curvature
are
\begin{eqnarray}
K^{r}_{\;r} & = & \frac{K}{3} + \frac{2C}{r^3}, \\
K^{\theta}_{\;\theta} & = & \frac{K}{3} - \frac{C}{r^3}, \\
K^{\phi}_{\;\phi} & = & \frac{K}{3} - \frac{C}{r^3}.
\end{eqnarray}
From these the $K^{ij}K_{ij}$ becomes
\begin{eqnarray}
K^{ij}K_{ij} = \frac{K^2}{3}+\frac{6C^2}{r^6}.
\end{eqnarray}
Inserting this into the equation for the lapse
\begin{eqnarray}
& & k\partial_{r}\partial_{r}\left(\delta k + k\int^{\infty}_{R}dr
\frac{\dot{C} - \frac{r^3}{3}\dot{K}}{r^2 k^3}\right) +
\left(\partial_{r}k + 2\frac{k}{r} \right)\partial_{r}\left(\delta k
+ k\int^{\infty}_{R}dr
\frac{\dot{C} - \frac{r^3}{3}\dot{K}}{r^2 k^3}\right) \notag \\
& & - \left(\frac{K^2}{3}+\frac{6C^2}{r^6}\right)\left(\delta k +
k\int^{\infty}_{R}dr \frac{\dot{C} - \frac{r^3}{3}\dot{K}}{r^2
k^3}\right) = \frac{\partial K}{\partial t}. \label{eqn:lapse
insertion}
\end{eqnarray}
The first term of the equation becomes
\begin{eqnarray}
k\partial_{r}\partial_{r}\left(\delta k + k\int^{\infty}_{R}dr
\frac{\dot{C} - \frac{r^3}{3}\dot{K}}{r^2 k^3}\right) & = &
k\left(\delta\partial_{rr}k + \partial_{rr}k\int^{\infty}_{R}dr
\frac{\dot{C} - \frac{r^3}{3}\dot{K}}{r^2k^3}\right. \notag \\
& & \left. - 2\partial_{r}k\left(\frac{\dot{C} -
\frac{r^3}{3}\dot{K}}{r^2k^3}\right) -
k\partial_{r}\left(\frac{\dot{C} -
\frac{r^3}{3}\dot{K}}{r^2k^3}\right)\right). \notag \\
\end{eqnarray}
Where the fundamental theorem of calculus has been used in order to
evaluate the derivative of the integral. Expanding the second term
\begin{eqnarray}
& & \hspace{-1 cm}\left(\partial_{r}k+ 2\frac{k}{r}
\right)\partial_{r}\left(\delta k + k\int^{\infty}_{R}dr
\frac{\dot{C} - \frac{r^3}{3}\dot{K}}{r^2 k^3}\right) =
\left(\partial_{r}k\ + 2\frac{k}{r} \right) \notag
\\
& & \hspace{3 cm}\times\left(\delta\partial_{r}k + \partial_r k
\left(\int^{\infty}_{R}dr \frac{\dot{C} - \frac{r^3}{3}\dot{K}}{r^2
k^3}\right) - k\left(\frac{\dot{C} - \frac{r^3}{3}\dot{K}}{r^2
k^3}\right)\right). \notag \\
\end{eqnarray}
Upon insertion of these expressions into (\ref{eqn:lapse insertion})
and factoring the result gives
\begin{eqnarray}
& & \hspace{-1 cm} \delta\left(k\partial_{rr}k + \left(\partial_r k
+ \frac{2k}{r}\right)\partial_{r}k - k\left(\frac{K^2}{3} +
k\frac{6C^2}{r^6}\right)\right) + \notag \\
& & \hspace{-1 cm} \int^{\infty}_{R}dr\frac{\dot{C} -
\frac{r^3}{3}\dot{K}}{r^2k^3}\left(k\partial_{r}k + \left(\partial_r
k + \frac{2k}{r}\right)\partial_{r}k - k\left(\frac{K^2}{3} +
k\frac{6C^2}{r^6}\right)\right) \notag \\
& & \hspace{-1 cm} -k^2\partial_{r}\left(\frac{\dot{C} -
\frac{r^3}{3}\dot{K}}{r^2k^3}\right) - \left(\partial_r k +
\frac{2k}{r}\right)\left(\frac{\dot{C} -
\frac{r^3}{3}\dot{K}}{r^2k^3}\right) - 2k\partial_{r}k
\left(\frac{\dot{C} - \frac{r^3}{3}\dot{K}}{r^2k^3}\right) =
\frac{\partial K}{\partial t}.
\notag \\
\end{eqnarray}
Evaluating the last line of the above equation gives
\begin{eqnarray}
& & \hspace{-1 cm}\delta\left(k\partial_{rr}k + \left(\partial_r k +
\frac{2k}{r}\right)\partial_{r}k - k\left(\frac{K^2}{3} +
\frac{6C^2}{r^6}\right)\right) + \notag \\
& & \hspace{-1 cm} \int^{\infty}_{R}dr\frac{\dot{C} -
\frac{r^3}{3}\dot{K}}{r^2k^3}\left(k\partial_{r}k + \left(\partial_r
k + \frac{2k}{r}\right)\partial_{r}k - k\left(\frac{K^2}{3} +
\frac{6C^2}{r^6}\right)\right) +
\frac{\dot{K}}{k} = \frac{\partial K}{\partial t}. \notag \\
\end{eqnarray}
Factoring the equation gives
\begin{eqnarray}
& & \hspace{-1 cm}\left(\partial_{rr}k + \frac{1}{k}\left(\partial_r
k + \frac{2k}{r}\right)\partial_{r}k - \left(\frac{K^2}{3} +
\frac{6C^2}{r^6}\right)\right) \notag \\
& & \hspace{2 cm}\left(\delta k + k\int^{\infty}_{R}dr\frac{\dot{C}
-
\frac{r^3}{3}\dot{K}}{r^2k^3}\right) + \frac{\dot{K}}{k}
= \frac{\partial K}{\partial t}. \notag \\
\end{eqnarray}
Which becomes
\begin{eqnarray}
\alpha\left(\partial_{rr}k + \frac{1}{k}\left(\partial_r k +
\frac{2k}{r}\right)\partial_{r}k - \left(\frac{K^2}{3} +
\frac{6C^2}{r^6}\right)\right) + \frac{\dot{K}}{k} = \frac{\partial
K}{\partial t}.
\end{eqnarray}
\end{widetext}
Upon evaluation of the term in the bracket we obtain
\begin{eqnarray}
\frac{Q^2}{r^4} + \frac{\dot{K}}{k} = \frac{\partial K}{\partial t}.
\end{eqnarray}
Here we see that the generalised lapse term for the
Reissner-Nordstr\"{o}m metric does not exactly solve
(\ref{eqn:elliptic cmc alpha}). However we see that the
$\frac{Q^2}{r^4}$ falls off very quickly as $O\left(r^{-4}\right)$
so will quickly become negligible and therefore far away from the
black hole will be equivalent to Schwarzschild.

\section{Conclusions}
The question of what constitutes a good or appropriate choice for the lapse and
shift has not been answered in this paper. There is no prescribed or systematic
method of determining how the choice will perform. However, from the calculations in the paper we have shown that there exists strong similarities in the behaviour, of the CMC slices, between the Reissner-Nordstr\"{o}m and Schwarzschild spacetimes. From figure \ref{fig:kvr} we see that the behaviour of $K$s are almost identical. As $Q$ is increased the graph of the Reissner-Nordstr\"{o}m $K$ is squeezed between $r_-$ and $r_+$ but the monotonically increasing behaviour is preserved.

Even though the generalised lapse function here did not satisfy
(\ref{eqn:elliptic cmc alpha}) fully we can take some insights from
this analysis. Again, after the calculations, once the $Q^2$ term is
set to zero, we recover the Schwarzschild regime. This suggests
there is validity in using the techniques developed for
Schwarzschild with problems involving the Reissner Nordstr\"{o}m
metric and by setting $Q^2 = 0$ checks and comparisons of the
numerical techniques on both spacetimes can be made.

\begin{acknowledgments}
Patrick Tuite would like to acknowledge IRCSET (Irish Research
Council for Science, Engineering and Technology) for their support
of this work.
\end{acknowledgments}

%\bibliography{CMCSlicingbib}
%\bibliographystyle{unsrt}
\end{document}